\documentclass{moriond}
\usepackage{cancel}

\newcommand{\Photo}{}

\begin{document}

\vspace*{4cm}
\title{SEARCH FOR $B^+ \to K^+ \nu \bar \nu$ DECAYS WITH AN INCLUSIVE TAGGING METHOD AT THE BELLE II EXPERIMENT}

\author{FILIPPO DATTOLA \\ on behalf of the Belle II Collaboration}

\address{Deutsches Elektronen-Synchrotron (DESY), Notkestrasse 85, 22607, Hamburg, Germany}

\maketitle

\abstracts{
This contribution illustrates a new search for the flavor-changing neutral-current decay ${B^+ \to K^+ \nu \bar \nu}$ performed by the Belle II experiment at the SuperKEKB asymmetric-energy electron-positron collider. In this study, a sample corresponding to an integrated luminosity of $63\, \rm{fb}^{-1}$ collected at the $\rm \Upsilon(4S)$ resonance and an additional sample of $9\, \rm{fb}^{-1}$ collected at an energy $60\, \rm MeV$ below the resonance are used. A novel technique, based on an inclusive tagging method and exploiting the topological features of the ${B^+ \to K^+ \nu \bar \nu}$ decay, is employed and it provides a higher signal efficiency with respect to the methods used in the previous searches. No significant signal is observed. An upper limit of $4.1 \times 10^{-5}$ is set on the ${B^+ \to K^+ \nu \bar \nu}$ branching fraction at the $90\, \%$ confidence level.}

\section{Introduction}
\begin{figure}[h]
\centering
\includegraphics[width=.35\textwidth]{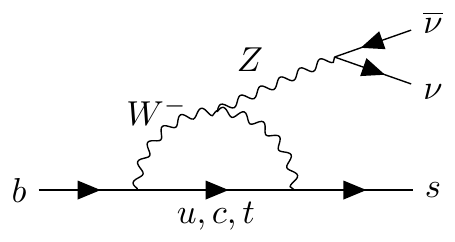}
\includegraphics[width=.35\textwidth]{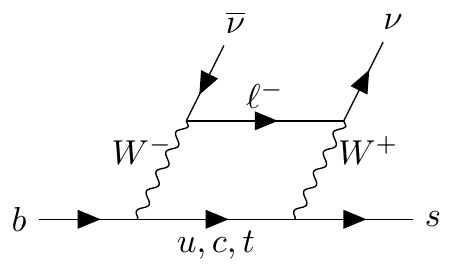}
\caption{Lowest-order SM Feynman diagrams for $b \to s \nu \bar \nu$ transitions.}
\label{fig:figure_1}
\end{figure}

In the Standard Model (SM), the ${B^+ \to K^+ \nu \bar \nu}$ decay belongs to the family of the ${b \to s \nu \bar \nu}$ flavor-changing neutral-current transitions. \footnote{Charge-conjugate channels are implied throughout this document unless explicitly stated otherwise.} Suppressed by the extended Glashow-Iliopoulos-Maiani mechanism \cite{Glashow:1970gm}, such processes can only occur at higher orders in SM perturbation theory via weak amplitudes involving the exchange of at least two gauge bosons, as illustrated in Fig. \ref{fig:figure_1}. In the specific case of the ${B^+ \to K^+ \nu \bar \nu}$ decay a branching fraction of $(4.6 \pm 0.5) \times 10^{-6}$ is predicted in the SM \cite{Blake:2016olu}. In all the analyses reported to date, no evidence of signal was observed and the current experimental upper limit on the branching fraction is estimated to be $1.6 \times 10^{-5}$ at the $90\, \%$ confidence level \cite{Zyla:2020zbs}.

In this study, a search for the $B^+ \to K^+ \nu \bar \nu$ decay is carried out with a novel \textit{inclusive tagging} approach, using the data from $e^+e^-$ collisions produced in 2019 and 2020 by the SuperKEKB collider \cite{Akai:2018mbz}, and a comprehensive description of the study is provided in Ref. \cite{Abudinen:2021emt}. The data correspond to $63\, \rm fb^{-1}$ integrated luminosity collected by the Belle II detector at a center-of-mass energy corresponding to the $\rm \Upsilon (4S)$ resonance, and containing approximately $68$ million $B \bar B$ pairs \cite{Aubert:2004pwa}, and to $9\, \rm fb^{-1}$ integrated luminosity collected at an energy $60\, \rm MeV$ below the resonance and used to constrain the yields of continuum processes (${e^+e^- \to q \bar q}$ with $q = u,d,s,c$ quarks and ${e^+e^- \to \tau^+ \tau^-}$).\\
The Belle II detector is composed of various subdetectors arranged in a cylindrical structure around the beam pipe: a pixel detector, fundamental for background rejection in this search, at a minimum radius of 1.4 cm; a silicon vertex detector and a central drift chamber; a time-of-propagation counter and an aerogel ring-imaging Cherenkov counter for charged-particle identification (PID); an electromagnetic calorimeter; a dedicated detector for $K^0_L$ and muon identification. A superconducting solenoid, operating at $1.5\, \rm T$, is situated outside the calorimeter. More details on the Belle II detector are in Ref. \cite{Abe:2010gxa}.
\section{The inclusive tagging}
\begin{figure}[h]
 \centering
 \begin{minipage}[t]{0.45\linewidth}
 \includegraphics[width=\textwidth]{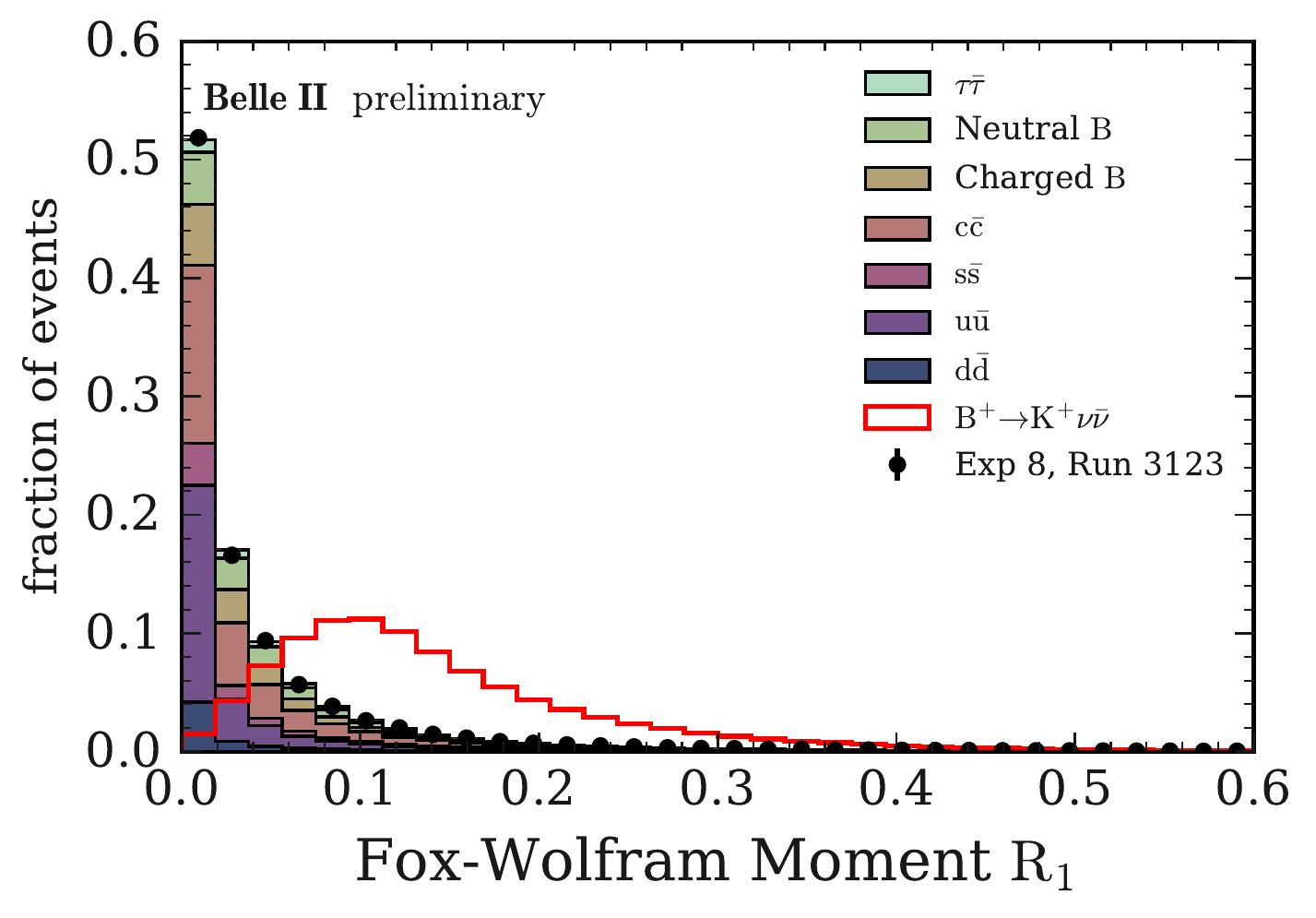}   
\caption{Distribution of the Fox-Wolfram variable $R_1$ used to train $\rm BDT_1$ and $\rm BDT_2$. The sum of backgrounds, signal, and data (from a single run for an illustrative comparison) are each normalized to unit area.}
   \label{fig:figure_2}
 \end{minipage}
 \quad
 \begin{minipage}[t]{0.45\linewidth}
   \includegraphics[width=1.1\textwidth]{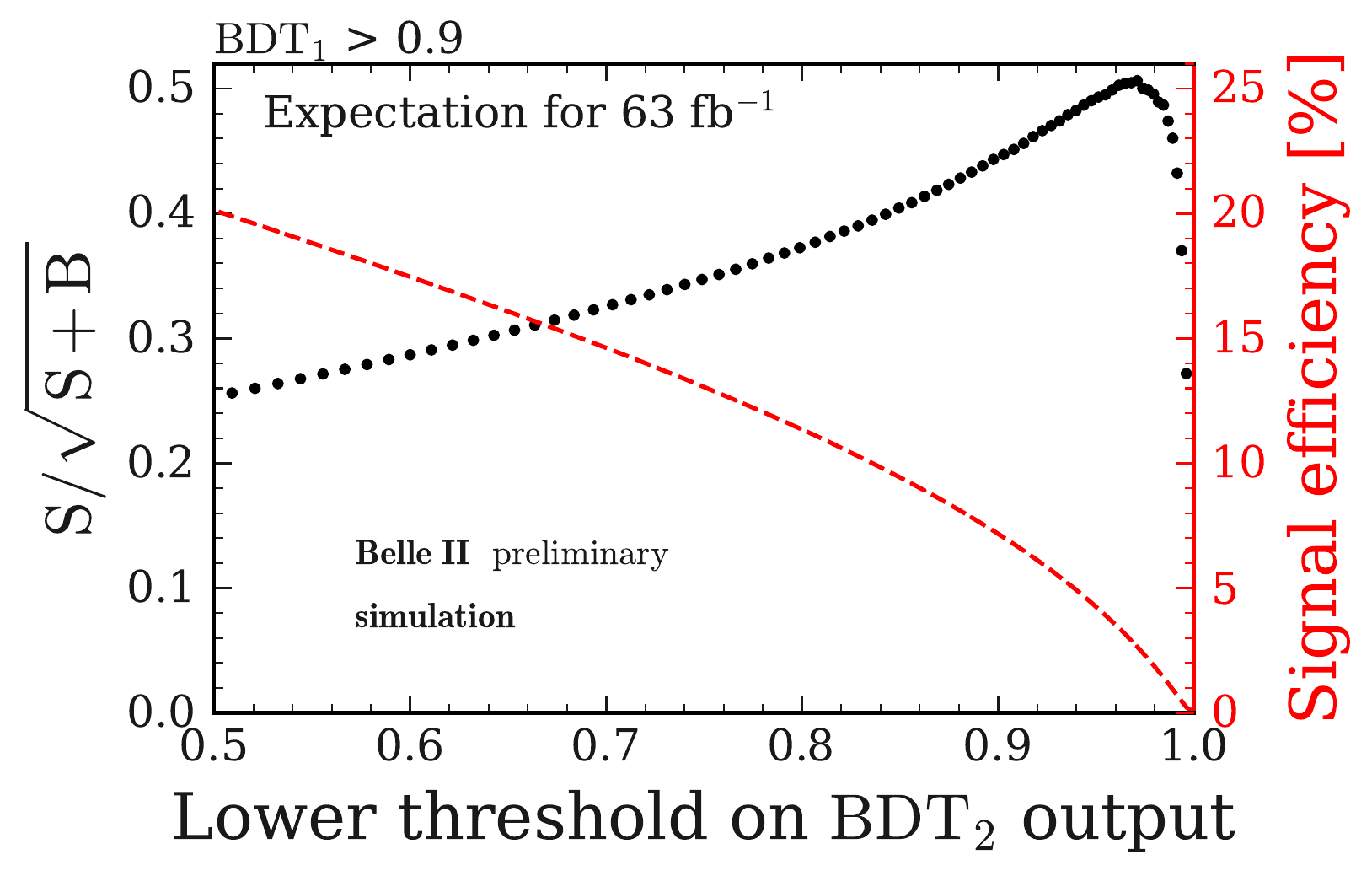}
    \caption{Signal efficiency (dashed line) and sensitivity $\rm S/\sqrt{S+B}$ (dots), where $\rm S(B)$ is the expected number of signal (background) events in $63\, \rm fb^{-1}$ integrated luminosity, as a function of the lower threshold on the $\rm BDT_2$ output.}
  \label{fig:figure_3}
\end{minipage}
\end{figure}

The previous searches for the ${B^+ \to K^+ \nu \bar \nu}$ decay adopted \textit{tagging} techniques where the accompanying $B$ meson in $e^+ e^- \to B \bar B$ events is explicitly reconstructed in a hadronic or a semileptonic decay \cite{PhysRevLett.86.2950,PhysRevD.87.111103,PhysRevD.87.112005,PhysRevD.82.112002,PhysRevD.96.091101}. Such methods suppress the contribution of background events, but at the cost of low signal-reconstruction efficiency, typically well below $1\, \%$.\\
The \textit{inclusive tagging} technique used in this search takes advantage of the specific topological features of the ${B^+ \to K^+ \nu \bar \nu}$ decay to distinguish it from the seven dominant background processes, namely,  generic decays of charged and neutral $B$ mesons as well as the five different continuum processes. In this study, the signal $K^+$ candidate in each event is chosen as the reconstructed charged-particle trajectory (track) carrying the largest transverse momentum and having at least one hit in the pixel detector.  The signal candidate is also required to satisfy PID requirements that suppress pion background. The remaining tracks are fit to a common vertex and, together with the remaining energy deposits, make up the \textit{rest of the event} (ROE). The specific features of the signal events are captured by several discriminating variables that are employed for signal identification.\\
In particular, a collection of 51 variables studied on simulated events is used to train two binary classifiers implementing the FastBDT algorithm \cite{FastBDTBelleII}. Such variables include general event-shape variables, variables describing the kinematic properties of the signal kaon candidate, variables related to the ROE, and variables that identify kaons from $D^0$ and $D^+$ decays  \cite{Bevan:2014iga}. The first classifier, $\rm BDT_1$, is trained using approximately $10^6$ simulated events of each of the seven background categories and on approximately $10^6$ simulated signal events. The most discriminating quantities are those describing the event shape, as the reduced Fox-Wolfram Moment $R_1$, shown in Fig. \ref{fig:figure_2}, and the modified Fox-Wolfram variables, which are related to the missing momentum in the event and to the momentum of the signal kaon candidate \cite{Fox:1978vw,PhysRevLett.91.261801}. To suppress background events with signal-like features, a second classifier, $ \rm BDT_2$, is introduced. It is trained with the same input variables used for $ \rm BDT_1$ computed on the events having $ \rm BDT_1 > 0.9$ out of $100\, \rm fb^{-1}$ integrated luminosity of simulated background and on $1.5 \times 10^6$ simulated signal events. This leads to a $10\, \%$ increase of the maximum signal sensitivity $\rm S / \sqrt{S + B}$, where $ \rm S(B)$ represents the expected number of signal (background) events, that goes up to approximately $50\%$ at $\rm BDT_2 > 0.95$, as shown in Fig. \ref{fig:figure_3}.\\
The signal region (SR) is defined at {$\rm BDT_1 > 0.9$}, {$\rm BDT_2 > 0.95$} and is divided into $3 \times 3$ bins in the ${\rm BDT_2 \times p_T}(K^+)$ space, with bins optimized to separate signal from background, by minimizing the expected upper limit on the $B^+ \to K^+ \nu \bar \nu$ branching fraction. The following three additional control regions (CRn, n=1, 2, 3) are selected in the ${\rm BDT_2 \times p_T}(K^+)$ space to constrain the background yields: CR1 consisting of $1 \times 3$ bins at $0.93 < \rm BDT_2 < 0.95$; CR2 and CR3 having the same bin boundaries as SR and CR1, respectively, but used for off-resonance data only. In the SR, the expected yields of the SM signal and of the backgrounds account for 14 and 844 events respectively, thus resulting in a signal efficiency of $4.3\, \%$. A study of the signal efficiency in the SR as a function of the dineutrino invariant-mass squared $q^2$ is illustrated in Fig. \ref{fig:figure_4}. 

\section{Analysis validation}
\begin{figure}[h]
 \centering
 \begin{minipage}[t]{0.45\linewidth}
\includegraphics[width=.99\linewidth]{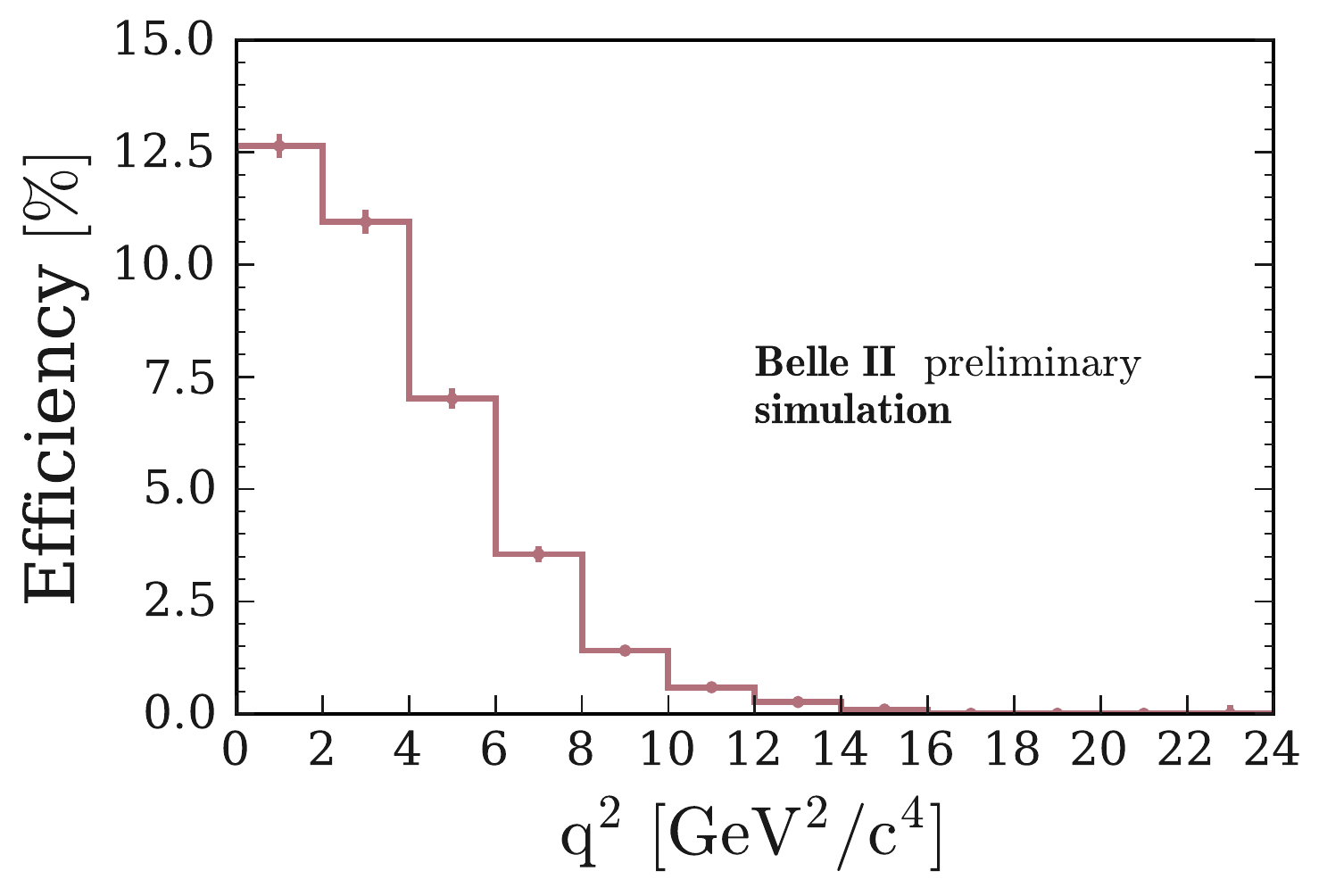}
\caption{Signal efficiency as a function of the dineutrino invariant-mass squared $q^2$ for events in the signal region (SR). Error bars correspond to the statistical uncertainty only.}
\label{fig:figure_4}
\end{minipage} 
\quad
  \begin{minipage}[t]{0.45\linewidth}
   \includegraphics[width=\textwidth]{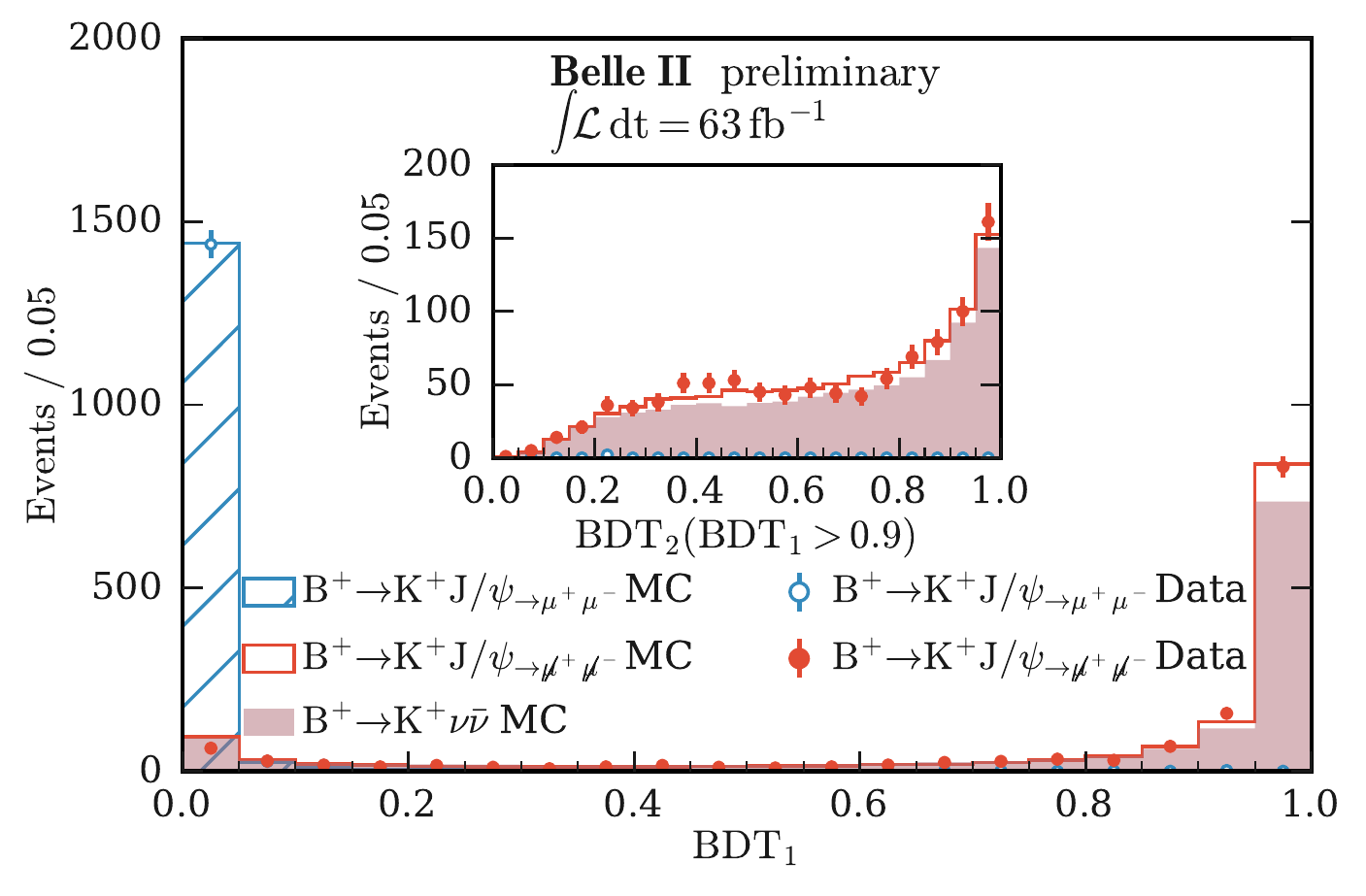}
    \caption{Distributions of the classifier outputs $\rm BDT_1$ (main figure) and $\rm BDT_2$ for $\rm BDT_1>0.9$ (inset), before (${B^+ \to K^+ J/\psi_{\to \mu^+ \mu^-}})$ and after (${B^+ \to K^+ J/\psi_{\to \cancel{\mu^+} \cancel{\mu^-}}}$) the muon removal and the update of the kaon-candidate momentum of selected ${B^+ \to K^+ J/\psi}$ events in simulation and data. The classifier outputs from simulated ${B^+ \to K^+ \nu \bar \nu}$ events are overlaid. The simulation histograms are scaled to the number of  ${B^+ \to K^+ J/\psi}$ events selected in data.}
  \label{fig:figure_5}
\end{minipage}
\end{figure}

The ${B^+ \to K^+ J/\psi}$ decay with ${J/\psi \to \mu^+ \mu^-}$ is used as an independent validation channel given its large branching fraction and distinctive experimental signature. ${B^+ \to K^+ J/\psi}$ events are selected in the $\rm \Upsilon(4S)$ data sample and in ${B^+ \to K^+ J/\psi_{\to \mu^+ \mu^-}}$ simulation by requiring the presence of dimuon pairs with invariant mass within $50\, {\rm MeV}/c^2$ of the known $J/\psi$ mass \cite{Zyla:2020zbs}.  The variables $| {\rm \Delta E }|  = | {\rm E}^*_B - \sqrt{\rm s}/2 |$ and ${\rm M_{bc}} = \sqrt{{\rm s}/(4c^4) - {{\rm p}^*_B}^2/c^2}$, where ${\rm E}^*_B$ and ${\rm p}^*_B$ correspond to the energy and the magnitude of the three-momentum of the signal $B$-meson candidate in the center-of-mass system of the beams, are required to be smaller than $110\, \rm MeV$ and larger than $5.25\, {\rm GeV}/c^2$, respectively. Such selection reduces the expected background contamination to $5\, \%$. Each event is then reconsidered as a ${B^+ \to K^+ \nu \bar \nu}$ event by ignoring the reconstructed muons from the $J/\psi$ decay and replacing the momentum of the kaon candidate with the generator-level momentum of the $K^+$ in a $B^+ \to K^+ \nu \bar \nu$ event from simulation. The results are illustrated in Fig. \ref{fig:figure_5}, where the distributions of the output values of $\rm BDT_1$ and $\rm BDT_2$ are shown. Good agreement between simulation and data is observed for the reconstructed events before (${B^+ \to K^+ J/\psi_{\to \mu^+ \mu^-}}$) and after (${B^+ \to K^+ J/\psi_{\to \cancel{\mu^+} \cancel{\mu^-}}}$) modifying the signal reconstruction.\\
Another validation test is performed by comparing the simulated continuum to the off-resonance data. Good agreement between data and simulation in the shape of the discriminant variable distribution is observed, as shown in Fig. \ref{fig:figure_6}, but a discrepancy between the data and simulation yields is present. Such discrepancy reaches $40\, \%$ in CR3 and CR4, motivating the introduction of a $50\%$ normalization uncertainty in the fit. The performance of $\rm BDT_1$ and $\rm BDT_2$ is also validated on events with $0.9 < \rm BDT_1 < 0.99$ and $\rm BDT_2 < 0.7$ , where the the $\rm \Upsilon(4S)$ on-resonance data and the corresponding simulation show good agreement.

\section{Statistical analysis and results}
\begin{figure}[ht]
 \centering
   \begin{minipage}[t]{0.45\linewidth}
 \includegraphics[width=0.98\textwidth]{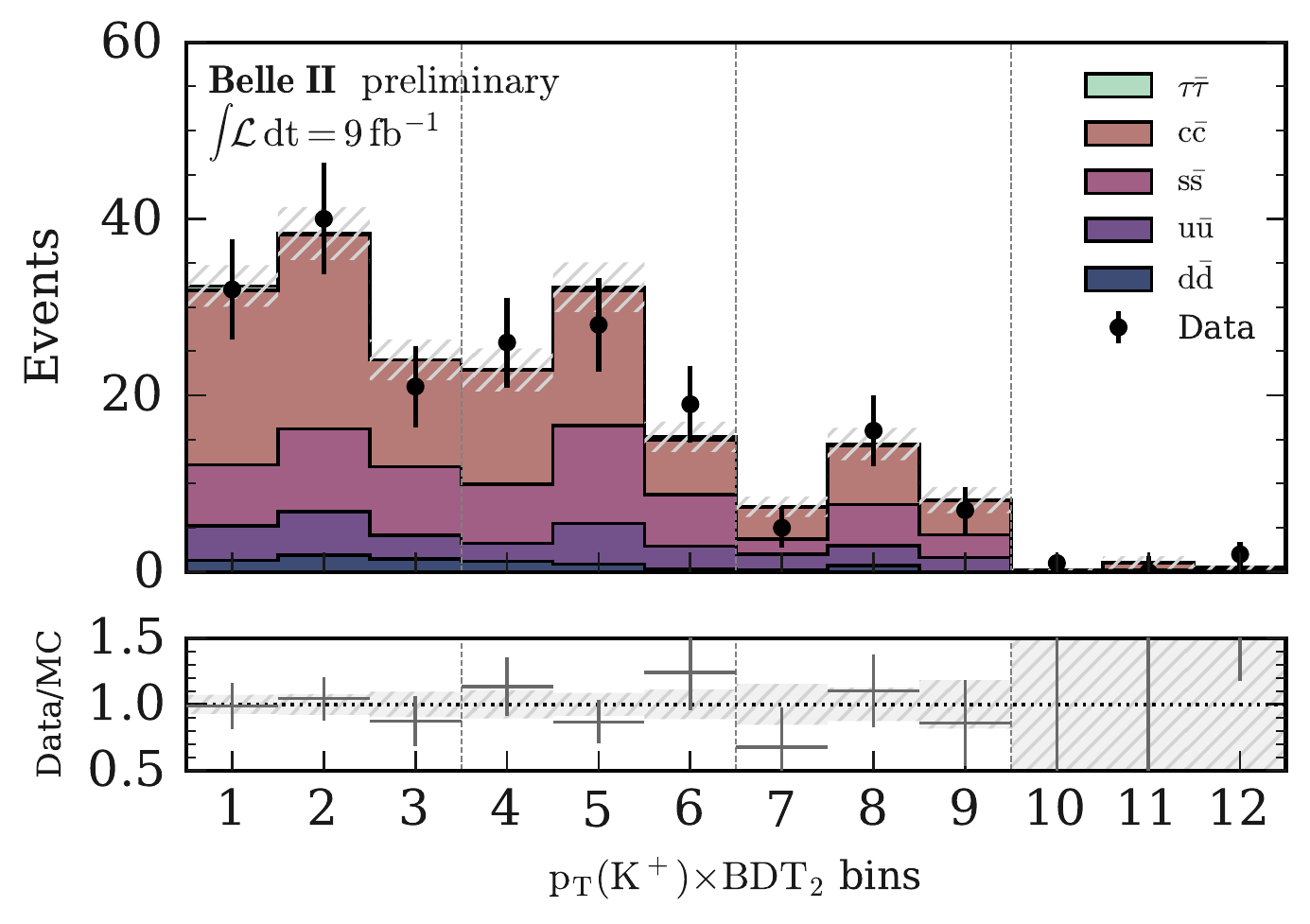} 
 \caption{Comparison of simulated events and continuum data in the two-dimensional control-region bins (the first three bins correspond to CR4, the latter nine bins to CR3). Yields in simulation are scaled to data by the data-to-simulation ratio ${\rm{data}/ \rm{simulation}=1.40\pm0.12}$, and the statistical uncertainty on the background yields is indicated by the hashed area.}
  \label{fig:figure_6} 
 \end{minipage}
\quad
  \begin{minipage}[t]{0.45\linewidth}
   \includegraphics[width=\textwidth]{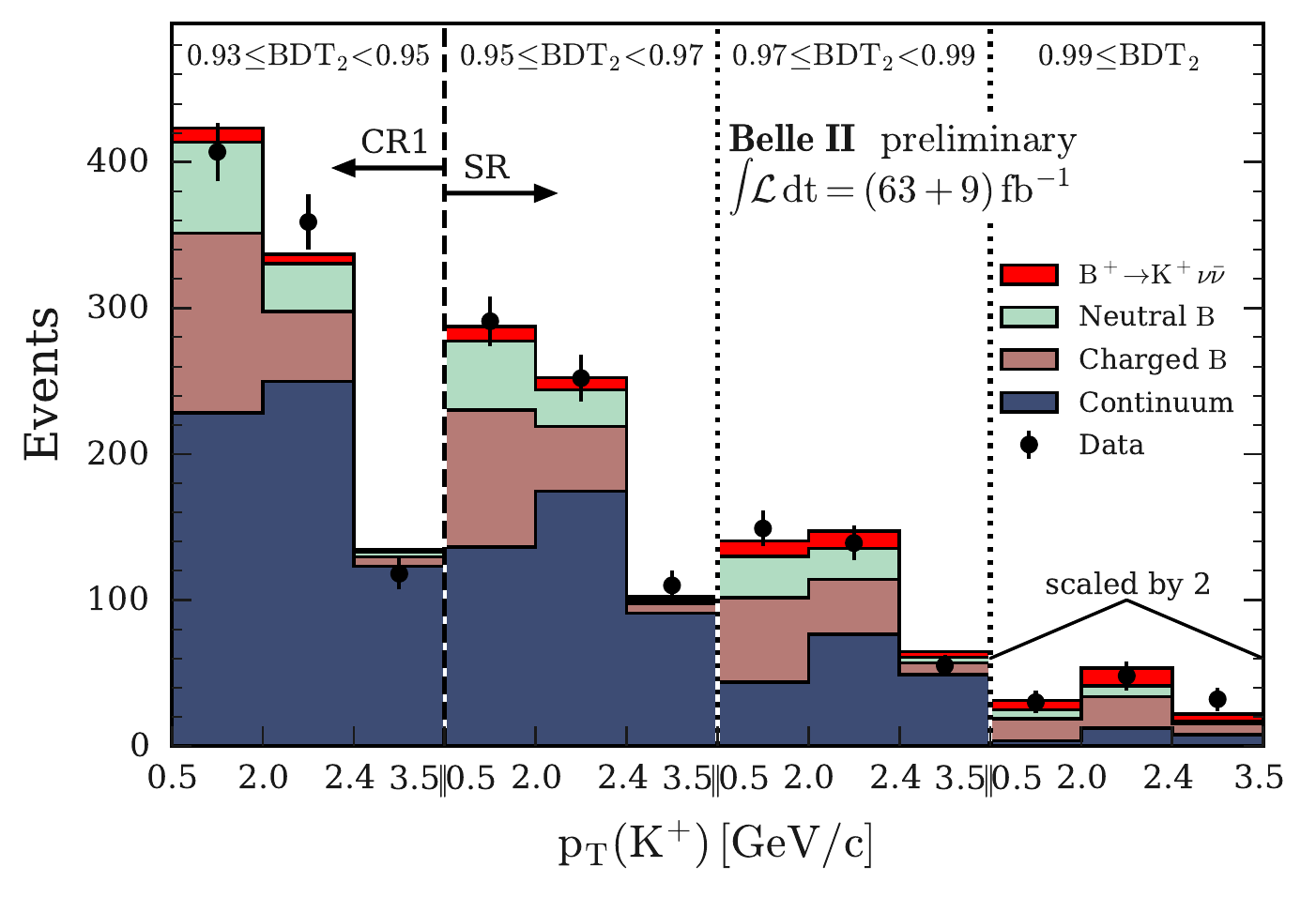}
    \caption{Yields in CR1 (first three bins) and SR (remaining nine bins) from on-resonance data and as predicted by the simultaneous fit to the on-resonance and off-resonance data, corresponding to an integrated luminosity of $63\, \rm{fb}^{-1}$ and $9\, \rm{fb}^{-1}$, respectively. The predicted yields are shown individually for the charged and neutral $B$-meson decays and for the sum of the five continuum backgrounds. All yields in the rightmost three bins are scaled by a factor of two.}
  \label{fig:figure_7}
\end{minipage}
\end{figure}

The statistical analysis is performed using the \texttt{pyhf} package  \cite{pyhf}. A binned likelihood is set up as the product of the Poisson probability density functions modelling the event counts in each of the 24 bins of the signal and control regions. The systematic uncertainties are introduced as nuisance parameters of the likelihood. The parameter of interest corresponds to the signal strength $\mu$, representing a multiplicative factor of the SM expectation. The major systematic uncertainty is related to the normalization of the background yields. The additional systematic uncertainties originate from the branching fractions of the leading $B$-meson decays, the PID correction, the SM form factors \cite{Buras:2014fpa}, the miscalibration of the hadronic and beam-background energy deposits in the calorimeter, the tracking inefficiency, and the limited size of the simulated samples. A comparison between the observations from data and the fit results in SR and CR1 is shown in Fig. \ref{fig:figure_7}.\\

The measured signal strength is ${\mu = 4.2^{+3.4}_{-3.2} = 4.2^{+2.9}_{-2.8}(\rm stat)^{+1.8}_{-1.6}(\rm syst)}$. The statistical uncertainty is determined by means of simplified simulated experiments corresponding to fluctuated observations in agreement with the Poisson statistics. The total uncertainty is obtained by a profile likelihood scan, where the fit is performed with $\mu$ fixed at values around the best fit value and the remaining parameters free. The systematic uncertainty is calculated by subtraction in quadrature of the statistical uncertainty from the total uncertainty. The result is translated into an observed branching ratio of ${[1.9^{+1.6}_{-1.5}] \times 10^{-5} = [1.9^{+1.3}_{-1.3}(\rm stat)^{+0.8}_{-0.7}(\rm syst)] \times 10^{-5}}$. No significant signal is observed and the expected and observed upper limits on the ${B^+ \to K^+ \nu \bar \nu}$ branching fraction are estimated using the $\rm CL_s$ method \cite{Read_2002}. Figure \ref{fig:figure_8} shows that at the $90\, \%$ confidence level the expected upper limit, derived in the background only hypothesis, is $2.3 \times 10^{-5}$ and the observed upper limit is $4.1 \times 10^{-5}$.

\section{Conclusion}
\begin{figure}[h]
\centering
  \begin{minipage}[t]{0.45\linewidth}
 \includegraphics[width=\textwidth]{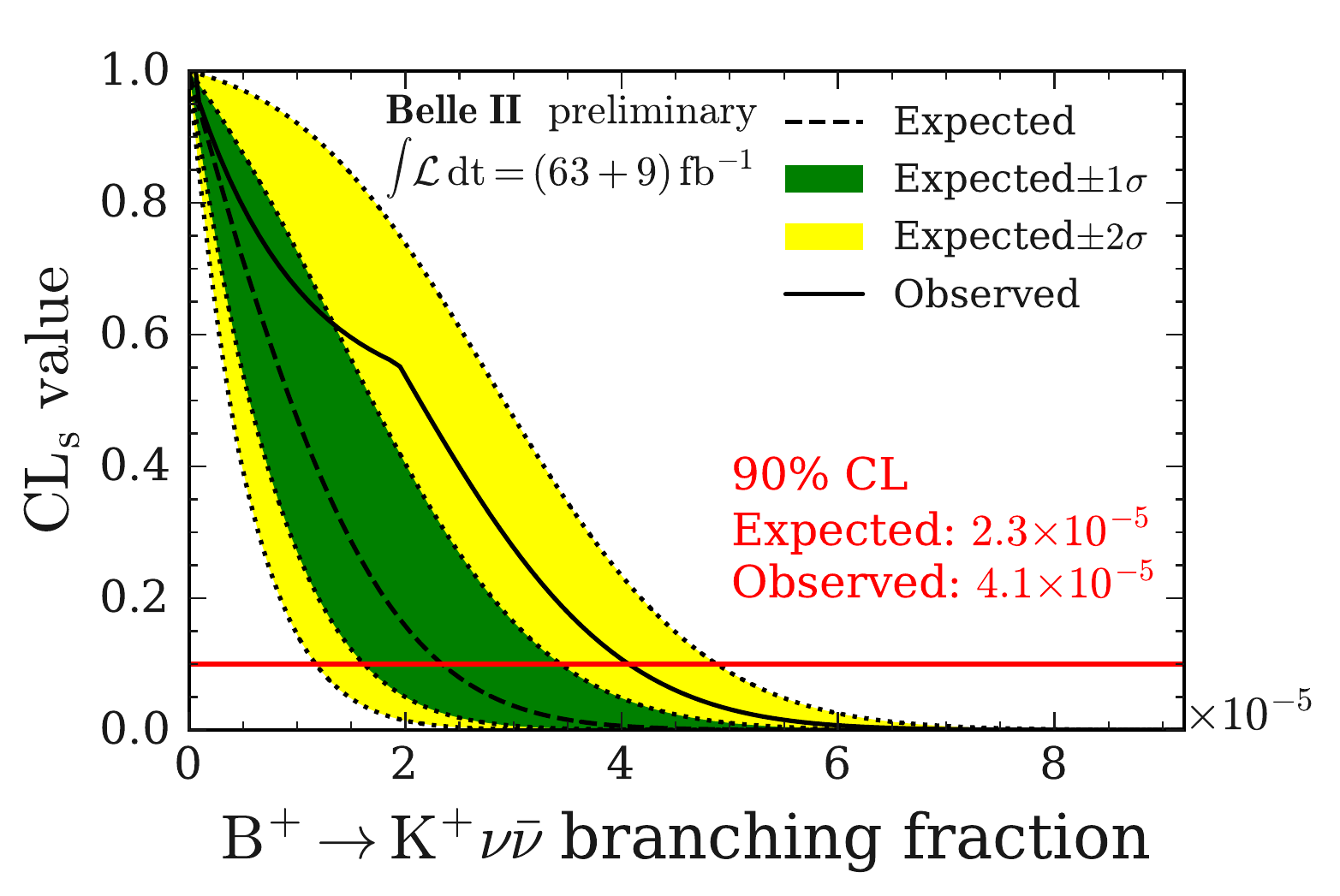} 
 \caption{$\rm CL_s$ value as a function of the branching fraction of ${B^+ \to K^+ \nu \bar \nu}$ for the expected and observed signal yields. In red the corresponding upper limits at the $90\, \%$ confidence level. The expected limit is derived for the background-only hypothesis.}
  \label{fig:figure_8} 
 \end{minipage}
 \quad
  \begin{minipage}[t]{0.45\linewidth}
\includegraphics[width=\textwidth]{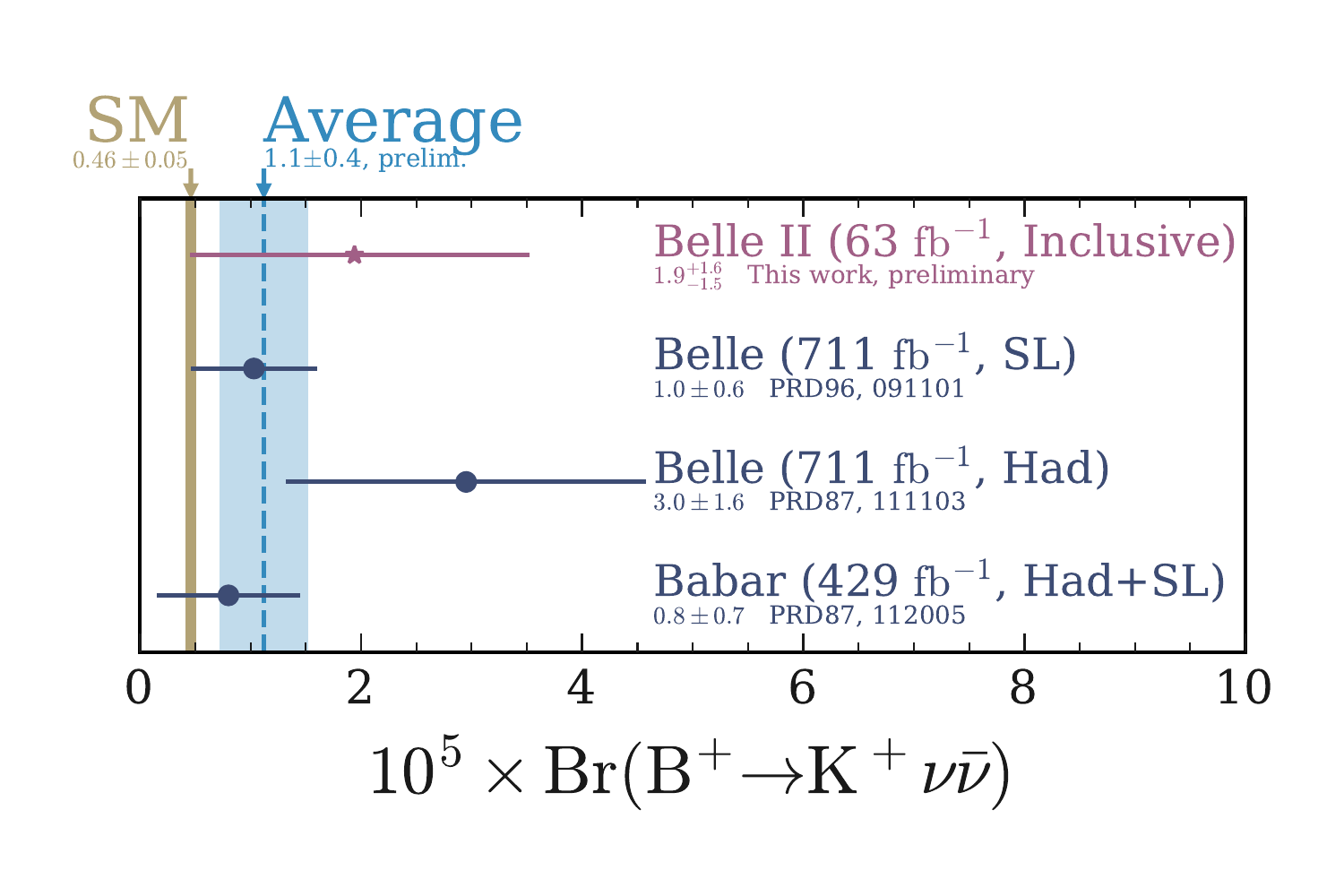}
\caption{Comparison of the branching fraction measured by Belle II and the previous experiments. The values reported for Belle are computed based on the quoted observed number of events and efficiency. The weighted average is computed assuming that uncertainties are uncorrelated.}
\label{fig:figure_9}
 \end{minipage}
\end{figure}

This contribution illustrates the first search for the ${B^+ \to K^+ \nu \bar \nu}$ decay with an inclusive tagging method. The study is performed on the data corresponding to $63\, \rm fb^{-1}$  integrated luminosity collected at the $\rm \Upsilon(4S)$ resonance by the Belle II detector, together with an additional sample of $9\, \rm fb^{-1}$ of off-resonance data. No statistically significant signal is observed and an upper limit of $4.1 \times 10^{-5}$ on the ${B^+ \to K^+ \nu \bar \nu}$ branching ratio is set at the $90\, \%$ confidence level. As illustrated in Fig. \ref{fig:figure_9}, the measurement is competitive with the previous searches, thus proving the capability of the inclusive tagging method.

\newpage
\section*{References}
\bibliography{references}

\end{document}